\documentclass[preprint,prl,showpacs]{revtex4-1}

\usepackage{times}
\usepackage{amsmath,amssymb}           
\usepackage{graphicx}                   
\usepackage{subfigure}                  
\usepackage{latexsym}
\usepackage{sidecap}

\begin{document}

\title{Dispersion in media containing resonant inclusions: where does it come from?}
\author{Fabrice Lemoult}
\author{Mathias Fink}
\author{Geoffroy Lerosey}
\email[]{geoffroy.lerosey@espci.fr}
\affiliation{Institut Langevin, ESPCI ParisTech \& CNRS UMR 7587, 10 rue Vauquelin, 75005 Paris, France}
\date{\today}

\begin{abstract}
We study the propagation of waves in a quasi 1D homogeneous host medium filled with various resonators. We first prove that a far field coupling between the elements explains its dispersive nature. This coupling is interpreted as a Fano interference between the incoming wave and the waves re-radiated by each resonator, which experience a phase shift at resonance. We propose a simple formalism that gives the complete dispersion relation of the medium in terms of the far field response of a single resonator. We prove that our approach applies to and unifies various domains such as metamaterials, hybridization band gap materials and designer's plasmons. Finally we show that those media, spatially random or organized on a scale larger than the wavelength, also present very interesting properties, which broadens the range of man made exotic materials.

\end{abstract}


\maketitle

Propagation media containing resonant inclusions have been studied for over 
a century in acoustics, electromagnetism or solid state physics. There exist 
some in nature, such as dielectrics, which contain enormous amounts of atoms. To calculate those materials 
permittivities one considers that each atom ``sees'' the same 
electromagnetic field and calculates the average field that 
takes into account an incoming wave as well as the overall response of the 
ensemble of atoms \cite{feynman}. This macroscopic view assumes that there are no 
variations of the electromagnetic field at the scale of the inter-atomic 
distance. Later on, the concept of polariton has been proposed to explain 
the propagation of electromagnetic waves in dielectrics 
\cite{Hopfield,Lagendijk}. The polariton approach of the index of refraction in 
dielectrics relies on mostly absorptive scattering and comes down to the 
original effective medium one \cite{Lagendijk}. 

There has been a renewed interest in systems consisting in arrays of 
resonators in a host medium since the last decade, notably owing to the 
emergence of metamaterials \cite{MetaBook1,MetaBook2}. 
Even though many metamaterials now rely on non resonant unit cells \cite{Pendry23062006,Leonhardt23062006}, we 
abusively ignore them here when mentioning 
metamaterials. John Pendry, who first proposed 
metamaterial presenting electric and magnetic activities in electromagnetism 
\cite{PendryWireMedium,pendrySRR}, adopted a macroscopic point of view and developed an 
approach based on field averaging to obtain the overall response of the resonators and
the effective properties of those man made materials \cite{MetaBook1,MetaBook2}. In this formalism
the period of the considered medium ought to be much smaller 
than the free space wavelength. Nowadays those 
materials are commonly studied from the far field which precludes any knowledge of the field 
at the unit cell level, and mostly to design media with negative index 
\cite{pendrySuperlens,Veselago,PRLSmith2000}. 

In the meantime, mainly in acoustics, there has been a large interest for 
materials presenting hybridization band gaps 
\cite{Liu2000,PhysRevB.84.094305,lee2009,PhysRevLett.100.194301,PhysRevB.65.064307,APL_Alice,PhysRevLett.77.2412,PhysRevB.75.235124}. The latter, contrary to conventional photonic or phononic crystals 
\cite{Yablonovitch,Page_phononique,Page2009,Joannopoulos}, present stop bands that are 
related to the resonant nature of the unit cell rather than the period of the medium. 
This phenomenon occurs when a resonator hybridises with the continuum of 
the plane waves of a homogeneous medium, giving rise to a binding and an 
anti-binding branches, that are separated by the so-called hybridization 
band gap \cite{PhysRevB.84.094305}. This effect is usually rather elusively justified by 
a level repulsion between the wave (the photon or the phonon), and a 
local resonance \cite{PhysRevB.84.094305}. 
Interestingly similar structures have been proposed 
to decouple antennas \cite{EBGmushroom,sieven} and are named 
electromagnetic band gap materials (EBG).

At this stage, one could wonder if we are not dealing with the same physical phenomenon albeit named differently. 
Metamaterials and hybridization based materials are indeed both obtained when 
a homogeneous medium is filled with resonant elements and they both present bands of permitted and prohibited propagation.
There are, however, drastic differences in 
the common understanding of those phenomena. Because of the deep subwavelength nature of metamaterials, their effective parameters and bandwidth are usually justified by a strong near field coupling between the elements  \cite{MetaBook1,MetaBook2}. On the contrary, hybridization effects suppose solely a coupling 
between free space waves and a single resonant element. One question
arises: are those phenomena identical, and if so, where does dispersion 
come from in those materials? 

In this letter, we use a very simplified model in order to grasp the physics 
of an array of resonators in a homogeneous medium. We first prove that, to a 
large extent, the response of a quasi 1D medium is governed by a far field 
coupling between the individual elements. This coupling can be understood as 
Fano interferences \cite{Fano,Miroshnichenko} between the incoming plane waves and 
the field reemitted by the resonators. We give a phenomenological 
description of this effect in terms of the frequency response of an 
oscillator. We propose a simple formalism that gives the dispersion relation of quasi 1D 
metamaterials using solely the far field transmission coefficient of a 
single unit cell and the period of the medium. We verify our approach and the obtained formalism on 
various media including random and wavelength scaled ones.

\begin{figure}
\begin{center}
\includegraphics[width=0.9\columnwidth]{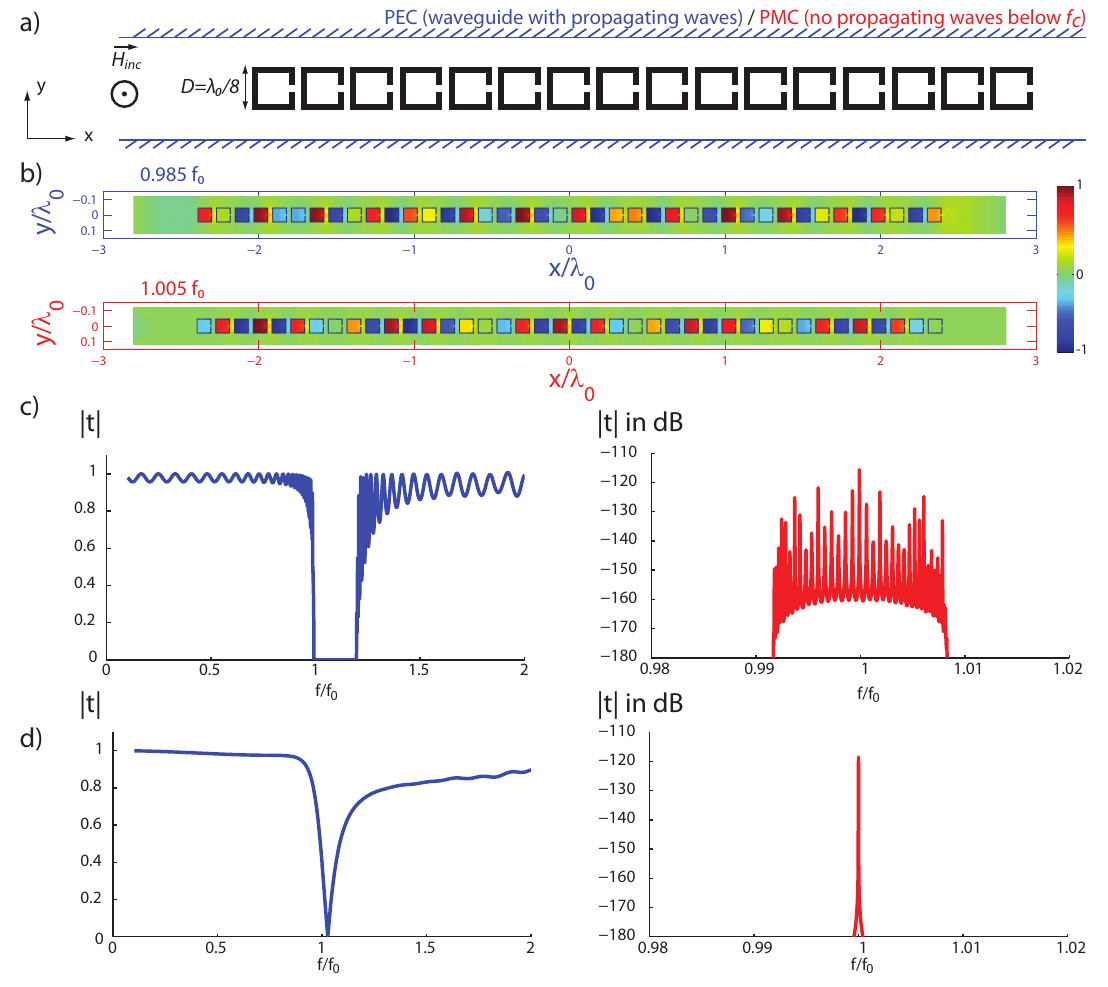}
\caption{\label{fig1} Simulated PEC (blue) or PMC (red) subwavelength waveguide filled with an array of SRR (a), maps of the fields for the two configurations and at two frequencies (b), transmission coefficients of the 40 SRR for PEC (blue) and PMC (red) case (c), transmission coefficients of a single SRR for PEC (blue) and PMC (red) case (d).}
\end{center}
\end{figure}

Our first goal is to identify the type of coupling that can give rise to an 
effective negative property in a metamaterial. To that aim, we study 
numerically a very simple system consisting in an array of 40 split ring resonators (SRR) made of 
perfect electric conductor (PEC) and placed in a waveguide of subwavelength width. 
The SRR resonate around $f_{0}$ for transverse 
magnetic polarized (TM) waves and have a square shape $\lambda 
_{\mathrm{0}}$/9 wide. Such a system is known to present a 
negative permeability. We perform the simulation in a 2D domain, and hence our approach is quasi 1D (Fig. 1.a.). 
Because we want to separate near field contribution from far field ones, we 
simulate the structure with two types of waveguides: the 
first has PEC boundaries and hence no cutoff for the impinging TM waves, 
while the second has perfect magnetic conductor (PMC) boundaries meaning 
that the impinging waves are evanescent along the waveguide. This way we can 
select which coupling we allow: near field only for 
PMC or near and far field couplings for PEC. In Figure 1.b, we map the magnetic field in the array of SRR 
for two different frequencies and for the PEC and PMC waveguides. Both
 show subwavelength varying patterns correspondent to a 
high effective index of refraction. More interesting are the plots of the 
transmission coefficients presented in Figure 1.c for the two studied cases. 
For the PMC waveguide, which forbids any far field type of coupling between 
the resonators, the modes of the system are symmetrically positioned 
around the resonance frequency $f_{0} $ of a single SRR on a narrow spectral 
range. This type of branch is typical of the 
tight-binding model commonly used in solid state physics \cite{kittel} or coupled resonators optical waveguides (CROWs) \cite{YarivCROW}, and cannot explain 
neither the effective negative permeability of the array of SRR, nor any band gap. On the contrary, the transmission coefficient of 
the PEC waveguide structure displays the wanted features: it is highly 
asymmetric, shows a wide band of propagating modes below $f_{0}$ and a wide stop band above. The band of 
propagating modes can be seen as a high permeability one, 
and the band gap as a negative 
permeability window. However another interpretation can be given.
Above the band gap, another 
set of modes seem to be observable from the transmission coefficient. This 
band, well known for materials presenting hybridization band gaps, is the 
anti-binding one, while that of the lower frequency and high wavenumber 
modes is the binding one. Those two branches are separated by a 
hybridization band gap. The response of this system can hence be explained 
by an hybridization between the resonance of a SRR and the continuum 
of free space waves.

From here two consequences appear. First, in order to obtain the expected properties of an array of SRR, one must take into account 
the far field coupling between the resonators. Second, this system, 
a 1D metamaterial, can be interpreted through the prism of 
hybridization. In order to obtain a deeper insight of the coupling involved, we simulate the response of a single 
resonator placed in the same waveguides. The transmission coefficients 
obtained are plotted in Figure 1.d. For the PMC case, the latter displays a 
symmetric shape consistent with a resonant 
tunneling of the electromagnetic energy at the resonance of the SRR. 
When the waveguide has PEC boundary conditions, the transmission coefficient 
has a singular profile since it presents a maximum of transmission right 
before the resonance and a minimum slightly after: this is typical of a Fano 
profile \cite{Fano,Miroshnichenko}.

\begin{figure}
\begin{center}
\includegraphics[width=0.85\columnwidth]{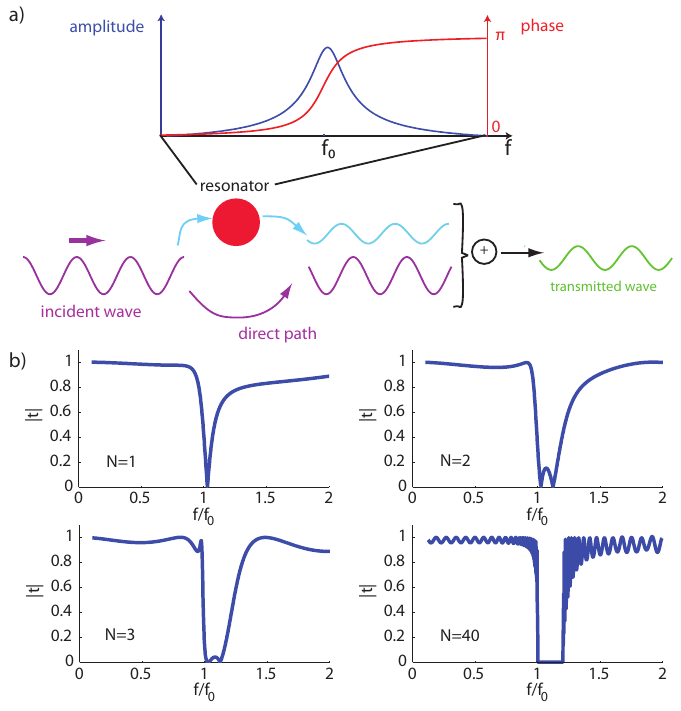}
\caption{\label{fig2} Scheme of the Fano interference in the PEC waveguide (a), transmission coefficients of 1,2,3 and 40 SRR  (b).}
\end{center}
\end{figure}

We prove now that Fano interferences explain the coupling between the 
resonators and its far field nature. The principle, schematised in 
Figure 2.a, can be explained in the following way. The resonator placed in 
the propagating waveguide scatters the incoming plane wave.
 Around resonance, the resonator accumulates energy and 
re-radiates waves which interfere with the unscattered incoming plane waves. 
Now the phase of the field in the oscillator changes abruptly at the 
resonance from 0 to $\pi$, meaning that 
its contribution can interfere constructively with the incoming plane waves 
or destructively. This explains the Fano profile and gives a very 
interesting information; since the resonator experiences a phase shift at 
the resonance, it can ``give'' some extra phase to the incoming plane wave, 
but it can also drastically suppress its transmission through destructive 
interferences \cite{Note}.

The principal features of the array of SRR can be inferred quite 
straightforwardly from this approach. On one hand, before resonance, the 
extra phase that originates from the resonators can compensate for the small 
phase shift accumulated through propagation over a unit cell, meaning that 
the field in the array can vary on scales as small as its period, hence 
leading to a high effective permeability. On the other hand, after 
resonance, the SRR respond in anti-phase and cancel the incoming plane wave 
which results in an hybridization band gap or equivalently an effective 
negative permeability.

Going from one resonator to an array, the effect of Fano type 
interferences are cumulative. In Figure 2.b, we plot the transmission 
coefficients of 1, 2, 3 and 40 SRR. 
The number of transmission maxima and minima equals the number of resonators on each branch. Each resonator gives rise to a Fano 
interference in the spectrum created by the previous one. This has two 
consequences; first, it widens the band of high or negative effective 
permeability. Second, a system of \textit{N} resonators supports 
\textit{2N} modes with distinct resonant frequencies, \textit{N} on the binding branch and \textit{N} on the anti-binding branch. Again, this differs from a tight-binding model, which gives only \textit{N} modes symmetrically positioned around the resonant frequency $f_{0}$. Naturally, an infinite system 
supports a continuum of modes resulting from a continuum of Fano 
interferences. Hence, the coupling between the resonators can be understood 
simply in terms of Fano interferences of far field components of the spatial 
spectrum.

We now utilize the formalism of the transfer matrix \cite{YarivBouquin} in order to obtain 
the dispersion relation of a linear array of resonators that are coupled 
through propagating waves only. To do 
so, we perform calculations very similar to those done on CROWs \cite{YarivCROW,YarivMatrices} and obtain a relation of 
the form \cite{Suppl}:

\begin{equation}
\cos\left( ka \right)=\Re \left( 
{\frac{1}{t}}e^{-j\frac{\omega 
}{c}a} \right)
\label{equ}
\end{equation}

where $k$ is the wavenumber in the medium, $a$ is the period of the medium and $t$ is 
the complex far field transmission coefficient of a single unit cell.

\begin{figure*}
\includegraphics[width=\columnwidth]{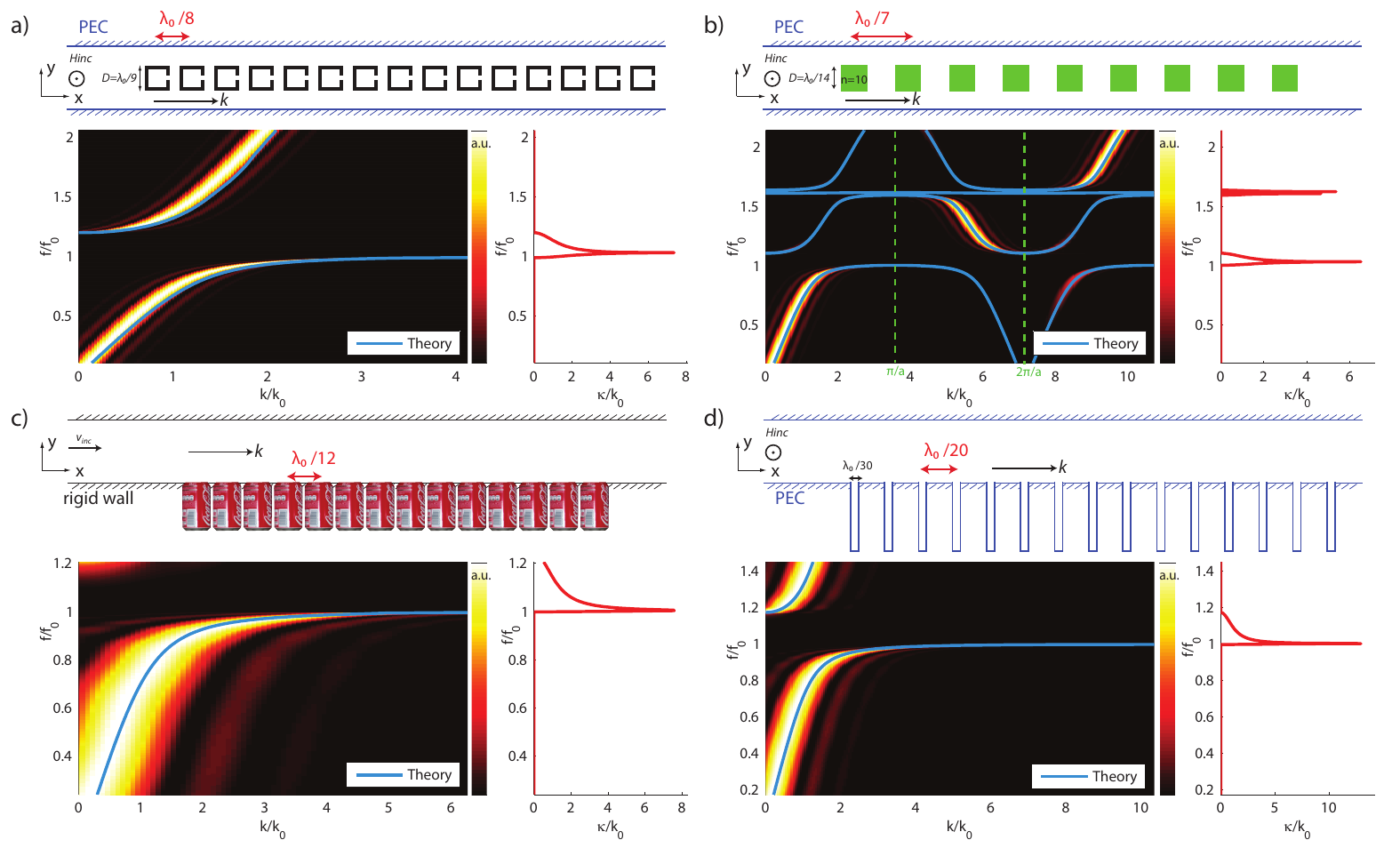}
\caption{\label{fig3} Structures and parameters, maps of the numerically obtained dispersion relations with superposed our theoretical predictions, and calculated attenuation coefficients for the: TM excited $\lambda_{\mathrm{0}}$/8 spaced array of SRR (a), TM excited $\lambda_{\mathrm{0}}$/7 spaced array of Mie resonators (b), acoustically excited $\lambda_{\mathrm{0}}$/12 spaced array of soda can or Helmholtz resonators (c), TM excited $\lambda_{\mathrm{0}}$/20 spaced array of PEC slits, that is, a designer's plasmons surface (d).}
\end{figure*}

This formula gives both the wavenumber $k$ of the propagating modes supported 
by an array of resonator, but also the attenuation coefficient $\kappa $ in 
the hybridization band gaps. We first verify its validity by numerically 
studying various types of resonators. Our simulations are lossless and all 
dimensions are given relatively to the wavelength at the first resonance of 
the considered unit cell $\lambda_{\mathrm{0}}$. We study the response to 
a plane wave of an array of 40 resonators in a waveguide. We 
simulate first a linear array of subwavelength SRR (Fig. \ref{fig3}.a). Then we apply 
our approach to Mie resonators, that is, squares of n$=$10 dielectric 
material (Fig. 3.b). We also simulate using acoustic waves the soda cans, 
that is the Helmholtz resonators, used in \cite{metalens_acoustique}, this time using a 3D waveguide (Fig. 3.c). 
Finally we study $\lambda_{\mathrm{0}}$/4 high, infinitely long slits in 
PEC, a building block of designer's plasmons 
\cite{PendryPlasmons,Vidal} (Fig. 3.d). We calculate the spatial spectrum of the electromagnetic or acoustic fields in those structures at each frequency and map the dispersion relations Figure 3.a-d. Meanwhile, we simulate each resonator independently in a 
waveguide and obtain its complex far 
field transmission coefficient. We superpose on the maps our theoretical dispersion relation and plot the calculated attenuation coefficients in the gaps using equation \ref{equ}.

Figure \ref{fig3} shows that for every medium under study, our simple 
formalism, based on propagating waves only, exactly reproduces the 
dispersion relation obtained by rigorously simulating the entire structure. This means that those structures, although organized on a deep subwavelength scale, are coupled mostly through propagating waves.
Furthermore, one each structure gives rise to a wide and efficient
hybridization band gap. For each medium, the frequency 
$f_{\mathrm{0}}$ separates the propagating modes and 
the hybridization band gap, which justifies the phenomenological explanation 
given above \cite{Note}. This means that all those quasi 1D materials have in common a 
unique physical basis: an hybridization between the resonator and the 
propagating waves. This approach permits to unify several distinct ones. For 
instance designer's plasmons \cite{PendryPlasmons,Vidal,Maier}, which are 
all based on a resonant unit cell, can be analysed under the same prism than 
negative permeability \cite{pendrySRR}, permittivity \cite{HuangPo04,Shapiro:06,doi:10.1088/0959-7174/7/2/006} or 
Young's modulus materials \cite{Liu2000,FangAcoustique,Fang2009}. This is in agreement 
with the fact that an interface between a negative and a positive effective 
property medium supports surface waves \cite{Shapiro:06,PhysRevB.75.195447} which we utilized 
in order to beat the diffraction limit from the 
far field \cite{metalens_acoustique,WRMLemoult1,WRMLemoult2,metalens}.

\begin{figure*}
\includegraphics[width=\columnwidth]{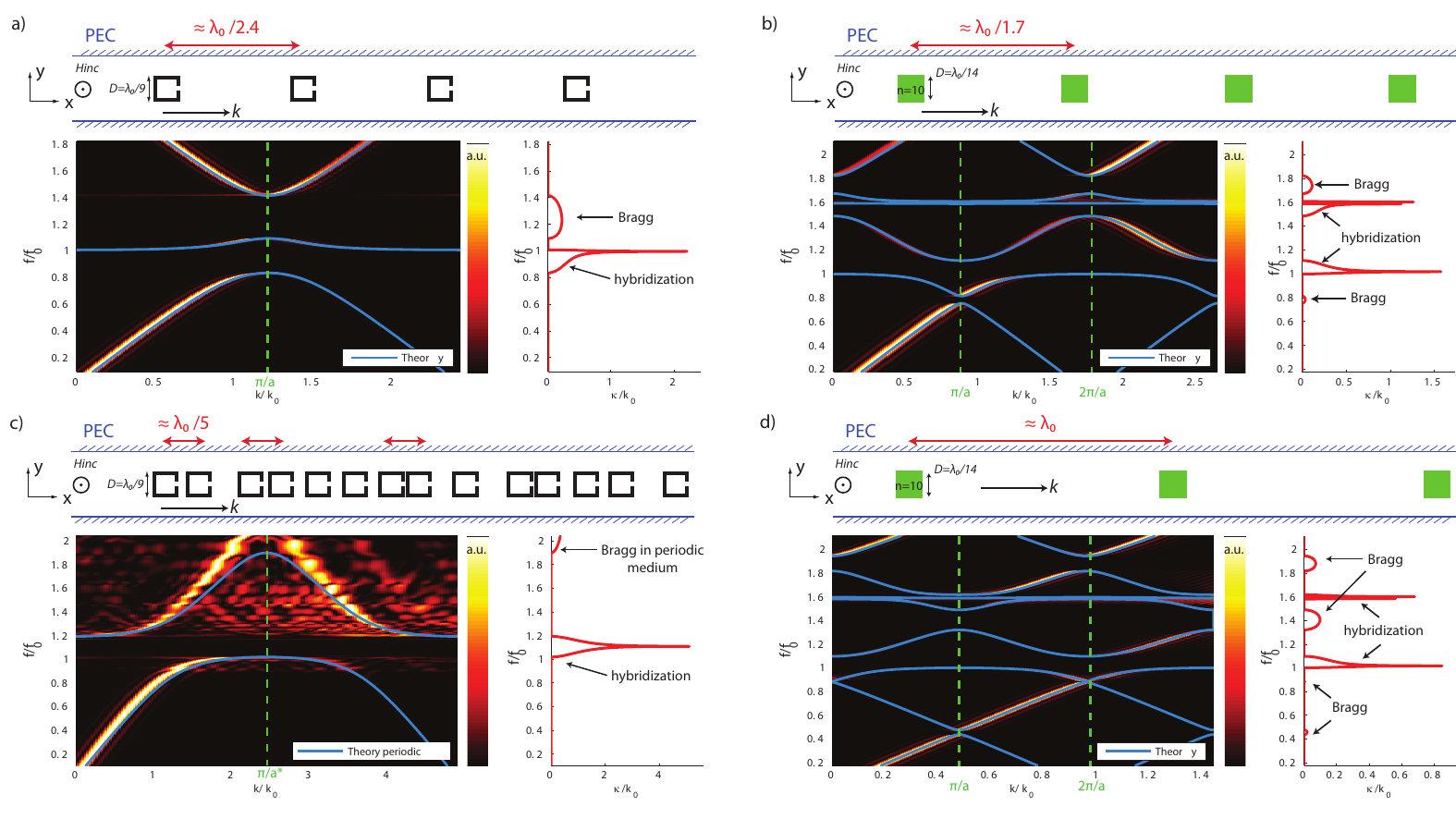}
\caption{\label{fig4} Structures and parameters, maps of the numerically obtained dispersion relations with superposed our theoretical predictions, and calculated attenuation coefficients for the: TM excited $\lambda_{\mathrm{0}}$/2.4 spaced array of SRR (a), TM excited $\lambda_{\mathrm{0}}$/1.7 spaced array of Mie resonators (b), TM excited and randomly positioned array of SRR (c), TM excited $\lambda_{\mathrm{0}}$ spaced array of Mie resonators (d).}
\end{figure*}

From these results, it is clear that near field couplings play a 
very little role in the dispersion observed in the systems studied here. 
This leads us to ask the following question: does one need to fulfill the 
effective medium condition of a deep subwavelength period in order to design 
media with interesting properties? To answer this question and demonstrate the possibilities offered by this Fano interference type of 
coupling, we present additional simulations. We want to 
demonstrate that hybridization band gaps, or equivalently bands of negative 
effective properties, exist even when the spacing between the resonators is 
increased and resist to disorder. To that aim we simulate four more systems. The first one consists in an array of SRR 
separated by $\lambda_{\mathrm{0}}$/2.4, a distance comparable 
to the period of typical Bragg crystals (Fig. \ref{fig4}.a). Similarly we study a $\lambda_{\mathrm{0}}$/1.7 periodic array of dielectric squares of index n$=$10 (Fig. 4.b). We simulate a chain of randomly positioned SRR
(average period a of $\lambda_{\mathrm{0}}$/5, Fig. 4.c). Finally we study an array of Mie resonators of period of $\lambda 
_{\mathrm{0}}$ (Fig. 4.d). Again we use equation \ref{equ} to superpose our theoretical 
dispersion relations and plot accordingly the 
attenuation coefficient in the band gaps. Those curves totally fit the simulations.

Figure 4 shows a very important feature common to all 
studied media: there exist in all the dispersion 
relations bands of propagation and hybridization band 
gaps regardless of the medium's order or period, thanks to the far field coupling between the resonators. Their positions, however, can be shifted due to the combined effects of interferences in between the unit cells and 
the resonant nature of the latter. Nevertheless, this proves that material 
made of resonant unit cells, even spaced by distances of the order or 
larger than the wavelength or random, can present negative effective properties.

Figure 4 also proves that Bragg bands and band gaps coexist with hybridization dilute systems, as was 
already proved in \cite{PhysRevB.84.094305,lee2009,PhysRevLett.100.194301,PhysRevB.65.064307,APL_Alice}, hence providing an even greater flexibility for designing materials with unusual properties. Due to their resonant nature, hybridization band gaps are much more efficient than Bragg ones. Mixing Bragg and hybridization effects 
could possibly lead to very interesting new phenomena. 

We note that disorder plays a different role for Bragg and hybridization types of bands. 
Indeed, in Figure 4.c, which shows the results of an array of SRR randomly 
positioned, the Bragg band gap has disappeared, while the hybridization one remains, a fact demonstrated in the past in acoustics \cite{PhysRevB.84.094305,lee2009,PhysRevLett.100.194301,PhysRevB.65.064307,APL_Alice}. This underlines the potential applications of hybridization band gaps in the visible.

This work finally appeals some concluding remarks. We have shown here that 
dispersion in arrays of resonators can be understood in terms of a far field 
type of coupling between the elements that arise from Fano interferences 
between the incoming waves and the waves re-radiated by the resonators. This 
approach gives a physical mechanism to the hybridization effect often 
referred to in acoustics and applies to metamaterials, locally resonant 
materials, designers plasmons and EBG structures. It demonstrates that wide 
bandwidth negative effective properties or hybridization gaps do not require any near field 
coupling and exist in random and wavelength scaled media.  
We believe that mixing Bragg and hybridization effects makes the physics of man made materials even 
richer and their properties more flexible. We underline however that some materials based on resonant unit cells can be 
also coupled through near field interactions, such as inductive or 
capacitive couplings in electromagnetics. This is typically the case for 
closely spaced SRR in 2D or 3D which can support magneto-inductive waves 
\cite{Shamonina}, or bubbles in acoustics \cite{APL_Alice,Valentin2}. Such near field coupling may 
be taken into account in our formalism, but this is out of the scope of this 
paper. Our approach requires a generalization to 2D and 3D materials. This could permit to design materials with exotic properties 
quite simply. Initial results on our experimental system of \cite{metalens_acoustique} support our intuition 
that 2D or 3D systems behave exactly as the 1D one studied here. This suggests that since the
Fano based dispersion in those media is linked to the forward scattering cross-section of the individual resonators, metamaterials are inherently non-local. 
Finally this idea should be 
transposable to all systems that support waves and contain resonant 
inclusions, which are very common in nature.

The  authors  acknowledge fundings from Orange Labs, and F. Lemoult acknowledges funding from French "Direction G\'en\'erale de l'Armement".

\bibliography{Draft}

\begin{thebibliography}{49}
\bibitem [{Not()}]{Note}%
  \BibitemOpen
  \href@noop {} {\bibinfo  {journal} {This depends on the scattering cross
  section of the resonator. For isotropic ones, the propagation band is before
  the resonant frequency and the gap after, for dipolar one it goes opposite. A
  more detailled study containing this analysis will be published elsewhere}\
  }\BibitemShut {NoStop}%
\end{thebibliography}

\end{document}